# Two-index model for characterizing site-specific night sky brightness patterns

Miroslav Kocifaj[1,2,*] and Salvador Bará[3]

[1]ICA, Slovak Academy of Sciences, Dúbravská cesta 9, 845 03 Bratislava, Slovakia.

[2]Department of Experimental Physics, Faculty of Mathematics, Physics, and Informatics, Comenius University, Mlynská dolina, 842 48 Bratislava, Slovak Republic.

[3]Departamento de Física Aplicada, Universidade de Santiago de Compostela, 15782 Santiago de Compostela, Galicia.

[*]*kocifaj@savba.sk*

**Abstract**

Determining the all-sky radiance distribution produced by artificial light sources is a computationally demanding task that generally requires an intensive calculation load. We develop in this work an analytic formulation that provides the all-sky radiance distribution produced by an artificial light source as an explicit and analytic function of the observation direction, depending on two single parameters that characterize the overall effects of the atmosphere. One of these parameters is related to the effective attenuation of the light beams, whereas the other accounts for the overall asymmetry of the combined scattering processes in molecules and aerosols. By means of this formulation a wide range of all-sky radiance distributions can be efficiently and accurately calculated in a short time. This substantial reduction in the number of required parameters, in comparison with other currently used approaches, is expected to facilitate the development of new applications in the field of light pollution research.

**Keywords:** Artificial sky brightness; Atmospheric optics; Scattering; Radiometry

1. **Introduction**

Modeling the all-sky radiance distribution produced by artificial light sources at arbitrary observing sites is an instrumental step for addressing several key issues in different fields of science and technology, including, among others, astronomical site characterization, intangible heritage preservation, and ecological studies (Walker 1970; Rich & Loncore 2006; Hölker et al. 2010; Falchi et al. 2011, 2016; Gaston et al. 2013, 2014; Kyba et al. 2015).

Quantitative estimations of this distribution can be obtained using radiative transfer models of different levels of complexity (Garstang 1989; Cinzano et al. 2001; Cinzano & Elvidge 2004; Kocifaj 2007, 2018a; Cinzano & Falchi 2012; Aubé 2015; Solano-Lamphar & Kocifaj 2016; Aubé & Simoneau 2018; Linares et al. 2018). For a comprehensive review, see Kocifaj (2016). A large set of factors determine the final result, including, among others, the spatial distribution, spectral power density, and angular radiant pattern of the artificial light sources (Kocifaj 2017, 2018b; Solano-Lamphar 2018), the state of the atmosphere (in particular, the type and concentration profiles of its aerosol constituents), as well as the spectral reflectance and geographical relief of the intervening terrain. A full site characterization shall ideally take into account the yearly statistics of the atmospheric conditions, including clouds (Solano-Lamphar & Kocifaj 2016), the time course of the artificial light emissions throughout the night (Dobler et al. 2015; Bará et al. 2018; Meier 2018), and the seasonal cycles of ground albedo due to changes in the vegetation and rainfall/snow (Puschnig et al. 2014; Coesfeld et al. 2018; Posch et al. 2018; Jechow & Hölker 2019). Given the extreme variability of several of these factors the calculations shall usually be done in a case by case basis.

Despite its great diversity, the all-sky radiance distribution cannot be completely arbitrary. Some basic physical processes constrain the structure of its possible shapes. On the one hand, single and multiple scattering usually give rise to a gentle and soft redistribution of artificial light in the sky, as seen from the observer. Excepting for extremely abrupt gradients at the boundaries of clouds or in the presence of obstacles along the observer's line of sight, the all-sky radiance distributions are expected to be smooth (i.e. class $C^{\infty}$) within the unobstructed domain of directions of the hemisphere located above the observer. On the other hand, for a layered atmosphere, the radiance distribution produced by an azimutally symmetrical source is expected to be symmetrical with respect to the vertical plane containing the zenith and the source, as seen from the origin of the observer's reference frame.

Different approaches can be used to describe the artificial radiance distribution across the celestial vault. The most immediate option is to specify the value of the radiance in all possible directions of observation (i.e., formally expressing it as a linear combination of directional Dirac-delta distributions), or, in practice, in a sufficiently dense discrete subset of them. However, this is clearly a suboptimal approach, because the artificial radiance is expected to be correlated in neighboring regions of the sky due to the smoothing produced by the scattering of light. The existence of these correlations allows to devise the possibility of obtaining more efficient mathematical descriptors. In this context, more efficient means requiring a substantially smaller number of parameters for successfully reconstructing the all-sky radiance distribution up to a given level of accuracy. Several well-known polynomial bases, like the Zernike or Legendre ones have been used to that end (Bará et al. 2014, 2015a,b). Expressing the sky radiance as a linear combination of the elements of these bases allows us to reduce the required number of parameters by about three orders of magnitude, from the ~$10^6$ data points typical of a pixel-wise description of an all-sky image to ~$10^3$ linear combination coefficients.

However, neither of these polynomial bases have originally been developed with the explicit aim of describing night sky brightness (NSB) distributions. It can be anticipated, then, that more efficient sets of functions could be found, that allow reconstructing the all-sky NSB using even less independent parameters. These new functions shall be built upon the available knowledge about the physical process that determine the structure of the artificial NSB. As a first step in this direction we develop in this work one such family of functions, that effectively fit the observed radiance distributions, and whose functional form is only dependent on two single parameters that can take continuous values within their respective domains of definition. These parameters are closely related to two basic physical properties of the atmosphere, the attenuation of light along the propagation paths and the overall asymmetry of the combined scattering processes. Each individual light source gives rise to an elementary NSB distribution described by these functions, and the overall NSB is easily calculated as a sum over sources.

The structure of this paper is as follows. In Section 2 we develop the proposed set of functions, as well as its rationale in terms of daylight distributions. In Section 3 we analyze two special cases of practical interest, that of a single dominant light source and that of purely isotropic scattering, respectively. In Section 4 we provide numerical results for a set of canonical cases that allow to get insight into the NSB behavior for different values of the overall attenuation and scattering asymmetry. In Section 5 we present observational validation results. Conclusions are summarized in Section 6.

## 2. A daylight analog for modeling the NSB distribution

Unlike the NSB, the angular distribution of daylight is due to scattering in the whole atmospheric column. The primary source of NSB is the artificial light from cities or towns, however, a common feature of both NSB and daylight are their atmospheric drivers. A photon undergoes scattering and attenuation on its path independent of the light source, so the physics of light propagation remains the same. This includes the transmission coefficients or scattering phase function as well. The key differences between daylight and NSB calculations are in optical path lengths and scattering geometries. The momentary sun position during sunset or sunrise is perhaps the best analogy to the NSB geometry with light sources visible at horizon. The sky radiance distribution, $L(z, A)$ then refers to Eq. (15) in Kocifaj and Kránicz (2011). This equation has been primarily developed to model the spectral radiance, but there are a number of evidences that radiance and luminance distributions share the same structure, meaning that both have measured and theoretical qualitative correspondences (Rossini & Krenzinger 2007). Due to this qualitative similarity a coefficient of proportionality is by far the most important source of differences between the radiometric and photometric distributions of the clear sky (Brunger & Hooper 1993). Therefore, for a model of adiabatic atmosphere in hydrostatic equilibrium and short optical paths in the lower atmosphere, the single-scattering radiance (or luminance) can be approximated as follows

$$L(z, A) = F_0 T_F\left(z, t, z_0 = \frac{\pi}{2}\right) P\left(z, A, A_0, z_0 = \frac{\pi}{2}\right), \qquad (1)$$

where $z$ and $A$ are the zenith and azimuth angles, respectively, of the direction of observation, and in our case the source is assumed to be on the horizon ($z_0 = \pi/2$), at an azimuth angle $A_0$. Whereas for daylight calculations $F_0$ is the extra-atmospheric spectral (or broadband) solar irradiance (or illuminance), for our present purpose of NSB calculations it will be a function characteristic of the source, with dimensions of irradiance (or illuminance), that can be experimentally determined as indicated below. We will use the common term 'radiance' throughout this paper to characterize $L$ unless otherwise indicated. $T_F$ introduced in Eq. (1) is a transmission function that carries information about the different atmospheric attenuation of beams propagating through different air masses, Eq. (2), being $t$ a scaling parameter with an obvious interpretation of optical thickness. $P$ is the scattering phase function. The above radiometric functions generally depend on wavelength, but this is not indicated for sake of brevity. For broadband radiance $T_F$, $t$, and $P$ need to be replaced by their spectrally-averaged equivalents. In addition, luminance computations require the spectral luminous efficiency for an individual observer to be used as a weighting function. Due to the boundary conditions and the linearity of the radiative transfer equation, the radiance ($L$) is directly proportional to the phase function (see e.g. Eq. 6.5.1b in Liou 2002 or Eqs. 4.23-4.25 in Hovenier et al. 2004). The transmission function introduced in Eq. (18) of Kocifaj & Kránicz (2011) is expressed here as follows

$$T_F\left(z, t, \frac{\pi}{2}\right) = \frac{M(z)}{M\left(\frac{\pi}{2}\right) - M(z)} \left[exp\{-M(z)t\} - exp\left\{-M\left(\frac{\pi}{2}\right)t\right\}\right], \qquad (2)$$

where the optical air mass $M(z)$ is computed in accordance with the formula

$$M(z) = \frac{2.0016}{\sin h + \sqrt{\sin^2 h + 0.003147}} \quad , \quad h = \frac{\pi}{2} - z, \qquad (3)$$

introduced by Gushchin (1988), however, in a bit different form. Therefore $M\left(\frac{\pi}{2}\right) \approx 35$. $T_F$ is a continuous, finite function of $z$ which, in the limiting case of $z$ approaching $\frac{\pi}{2}$, takes the value

$$T_F\left(z, t, \frac{\pi}{2}\right) = M\left(\frac{\pi}{2}\right) t\, e^{-M\left(\frac{\pi}{2}\right)t} \cong 35\, t\, e^{-35t}. \qquad (4)$$

In a cloud-free atmosphere the phase function ($P$) is derived from the aerosol and Rayleigh scattering functions. The weighted contribution of both can be fitted easily by some $g$-valued Henyey-Greenstein function

$$P\left(z, A, A_0, \frac{\pi}{2}\right) = \frac{1}{4\pi} \frac{1 - g^2}{[1 + g^2 - 2g\cos\theta]^{3/2}}, \qquad (5)$$

where the scattering angle $\theta$ can be determined from spherical trigonometry. Because $z_0 = \frac{\pi}{2}$ we have

$$\cos\theta = \cos z \cos z_0 + \sin z \sin z_0 \cos(A - A_0) = \sin z\, \cos(A - A_0). \qquad (6)$$

The function $F_0$ associated with the light source is unknown, in a vast majority of cases, but it can be either determined experimentally or found linearly proportional to its total lumen output. It follows from Eqs. (1-6) that the radiance of the beam propagating along the line-of-sight from the light source and measured at the entrance of the observer's detector is

$$L\left(\frac{\pi}{2}, A_0\right) = \frac{F_0}{4\pi} \frac{1+g}{(1-g)^2} M\left(\frac{\pi}{2}\right) t\, e^{-M\left(\frac{\pi}{2}\right)t}, \tag{7}$$

therefore the value of $F_0$ corresponding to the $i$-th source of light is calculated as follows

$$F_{0i} = L_i\left(\frac{\pi}{2}, A_{0i}\right) \frac{4\pi}{M\left(\frac{\pi}{2}\right)t} \frac{(1-g)^2}{1+g}\, e^{M\left(\frac{\pi}{2}\right)t}, \tag{8}$$

where $L_i\left(\frac{\pi}{2}, A_{0i}\right)$ is the line-of-sight radiance, measured at the detector, of the $i$-th city or town located on the horizon and whose azimuth angle is $A_{0i}$ in the observer's reference frame. The total night sky brightness due to $N$ azimuthally separated light sources surrounding the observer is then

$$\begin{aligned} L(z, A) = &\frac{(1-g)^2}{1+g} \frac{M(z)}{M\left(\frac{\pi}{2}\right)t} \frac{e^{\left[M\left(\frac{\pi}{2}\right)-M(z)\right]t} - 1}{M\left(\frac{\pi}{2}\right)-M(z)} \times \\ &\times \sum_{i=1}^{N} \frac{L_i\left(\frac{\pi}{2}, A_{0i}\right)(1-g^2)}{[1+g^2 - 2g\sin z \cos(A - A_{0i})]^{3/2}}. \end{aligned} \tag{9}$$

The radiances $L_i$ and azimuths $A_{0i}$ characterize the light sources as seen from the observer location: $L_i$ are obtained experimentally on site or inferred from satellite radiance data (LPinfo 2019; Elvidge et al, 2017); the azimuth angles are easy to derive from geographical maps taking into account the position of measuring site. The only unknown parameters left are $g$ and $t$. They are spectral or spectrally-weighted quantities depending on whether the measurements are made by using a narrow-band filter (or spectrometer) or a broad-band radiometer (or photometer). The main strength of the model we have developed is the extremely low number of free parameters – in fact, two scaling parameters are sufficient to model the all-sky radiance distribution. It has to be emphasized that conventional databases of night sky brightness measurements are to archive SQM data, which is basically 1:1 relation, i.e. one scalar value is to characterize NSB in one direction (typically in zenith).

## 3. Two special cases

The formula for the radiance in Eq. (9) is intended to be used in two ways: a) to model and/or predict the sky state in an arbitrary location from the two physical parameters, specifically $g$ and $t$ (see e.g. demonstrations in Figs. 4-6), and b) to determine these scaling parameters from the observed NSB. As the parameter $t$ increases as the atmosphere becomes more turbid, the intensity of a light beam significantly decays. Therefore high values of $t$ can be representative for heavily polluted cities or metropolitan areas, while low values of $t$ are rather typical for clean atmosphere, especially in mountain regions. The smaller the particle size, the smaller $g$ we can expect. Normally the asymmetry parameter increases as the particle size approaches large scales (Wickramasinghe 1967; Moosmüller & Ogren 2017), so the large particles produced from agricultural or construction activities are characterized by preferably forward-lobed scattering functions. Of course, the classical asymmetry parameter and the $g$ used in our model are not identical. However, they both share the same features.

The case (b) discussed above often transforms into an optimization problem, i.e. to find the theoretical NSB distribution that best represents the experimental data. It is expected that a set of observations can produce more stable solutions for $g$ and $t$, but a single experiment should in principle be adequate to retrieve the scaling parameters. Two special regimes are investigated below, for which an analytical solution to Eq. (9) exists.

### 3.1. One light source dominating all others

In a heterogeneous territory the NSB characterization often reduces to the problem in which the light emissions from a single source dominate all others. However, such source not necessarily is a large human settlement, especially if seen at large distance from an observer. By far the largest NSB signal can be due to a small city in the vicinity of the measuring station. The zenith-normalized brightness in the great circle perpendicular to the vertical plane that contains the zenith and the azimuthal position of the light source is

$$\frac{L\left(z, A = A_0 + \frac{\pi}{2}\right)}{L(0,0)} = M(z) \left[\frac{e^{M\left(\frac{\pi}{2}\right)t - M(z)t} - 1}{e^{-t + M\left(\frac{\pi}{2}\right)t} - 1}\right] \left[\frac{M\left(\frac{\pi}{2}\right) - 1}{M\left(\frac{\pi}{2}\right) - M(z)}\right], \qquad (10)$$

which is the formula that allows for a straightforward determination of the parameter $t$ using radiance data taken at a few zenith angles. Fig. 1 demonstrates that Eq. (10) can simulate both increasing and decreasing trend of NSB when zenith angle goes from 0 to $\pi/2$. The low values of $t$ are to model a low turbid atmosphere, while $t$ tends to increase for shorter visual ranges (normally under high levels of air pollution). Fig. 1 shows that $t$ acts as a shaping factor for NSB when data analysis is confined to the above mentioned specific great circle. This is why $t$ can be easily assessed by drawing computed values along with experimentally determined zenith-normalized NSB data.

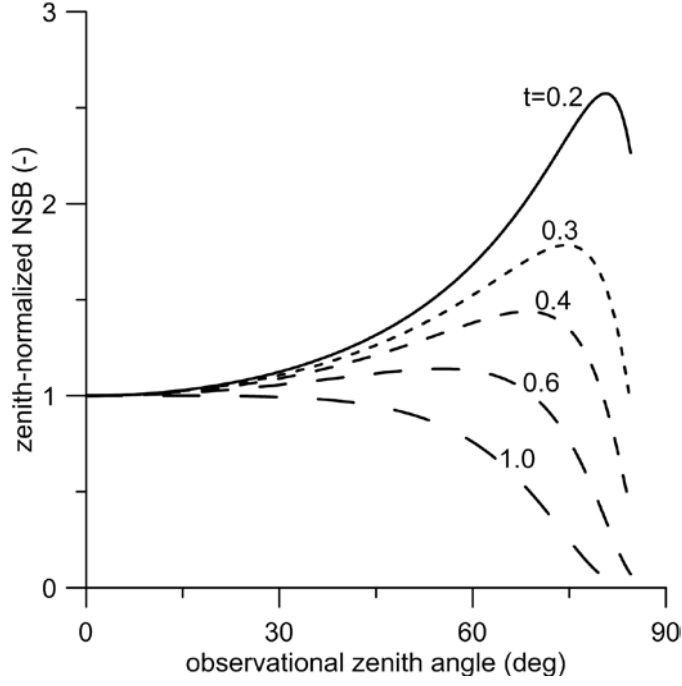

Fig. 1: Theoretical zenith-normalized night sky brightness in the plane perpendicular to the vertical plane passing through the azimuthal position of a light source and crossing the upper hemisphere in zenith. The computations are made using Eq. (10) for different values of $t$.

Assuming the quadrant of the sky opposite to that of the primary (dominant) light source is dark and no other sources group around it, the theoretical NSB at the horizon at $A = A_0 + \pi$ is

$$\frac{L\left(\frac{\pi}{2}, A_0 + \pi\right)}{L(0,0)} = \left(\frac{\sqrt{1+g^2}}{1+g}\right)^3 M\left(\frac{\pi}{2}\right) t \left[\frac{M\left(\frac{\pi}{2}\right) - 1}{e^{-t+M\left(\frac{\pi}{2}\right)t} - 1}\right] \quad (11)$$

which along with Eq. (10) comprises a complete and unique solution to the model, because $g$ can be obtained from the near-horizon value of NSB once the parameter $t$ is known. It is shown in Fig. 2 that the values of $L\left(\frac{\pi}{2}, A_0 + \pi\right)$ relative to $L(0,0)$ are a monotonic function of $g$, so the computation of the parameter $g$ from Eq. (11) is a straightforward procedure. Nevertheless, the clear sky radiance can only rarely be measured near the horizon because of obstacles or enhanced sky masking effect by distant clouds. Including as much unmasked sky elements to the minimization routine as possible is by far the best way to retrieve $t$ and $g$. The numerical processing then aims to minimize the differences between the theoretical and the experimental zenith-normalized NSBs. Employing the formalism of Eq. (9) the predicted NSB ratio due to $N$ light sources is

$$\frac{L(z,A)}{L(0,0)} = \frac{(1+g^2)^{\frac{3}{2}} M(z)}{\sum_{i=1}^{N} L_i\left(\frac{\pi}{2}, A_{0i}\right)} \left[\frac{e^{M\left(\frac{\pi}{2}\right)t - M(z)t} - 1}{e^{-t + M\left(\frac{\pi}{2}\right)t} - 1}\right] \left[\frac{M\left(\frac{\pi}{2}\right) - 1}{M\left(\frac{\pi}{2}\right) - M(z)}\right] \times$$
$$\times \sum_{i=1}^{N} \frac{L_i\left(\frac{\pi}{2}, A_{0i}\right)}{[1 + g^2 - 2g \sin z \cos(A - A_{0i})]^{3/2}},$$
(12)

which is the formula used in the numerical demonstrations (Section 4) and the experimental validation of the model by retrieving the parameters that best match the observed NSB (Section 5).

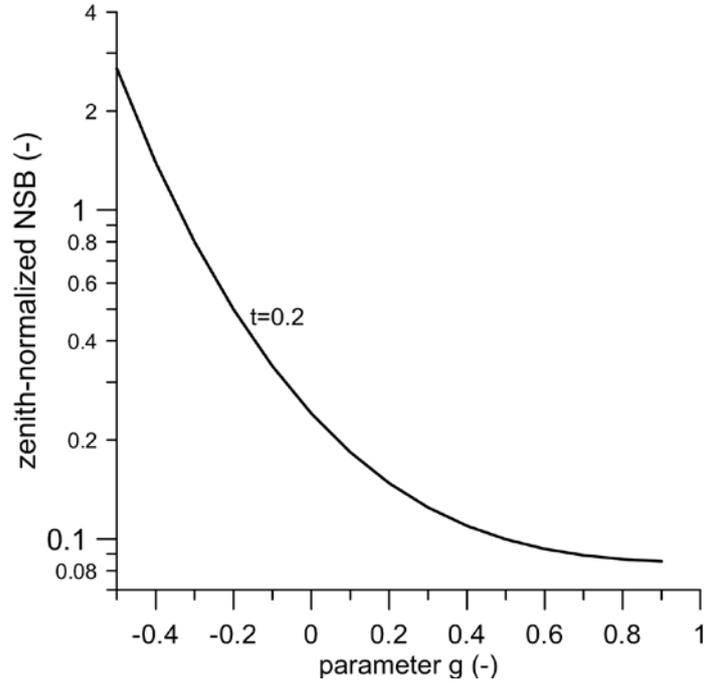

Fig. 2: Theoretical zenith-normalized NSB at the horizon opposite to the position of a light source. The parameter $t$=0.2, while $g$ is varied from -0.5 to 0.9.

### 3.2. Pure isotropic scattering

An isotropic scattering medium is considered to be an idealization of real physical embodiments of natural systems, however its theoretical treatment can have important consequences as it allows us to validate critical limits of the model, meaning that we can test and improve upon the theories. For isotropic media non-selective scattering is a common property, and the optical signal detected usually has undergone many scattering events. Therefore, the photon tracking on its path through that media is more-or-less impossible. The beams of light escaping such medium have non-preferred directions, thus supporting azimuthally uniform light fields (see Fig. 3). There are not too many systems in nature that satisfy the above conditions. A few of them are thick clouds or a dense fog.

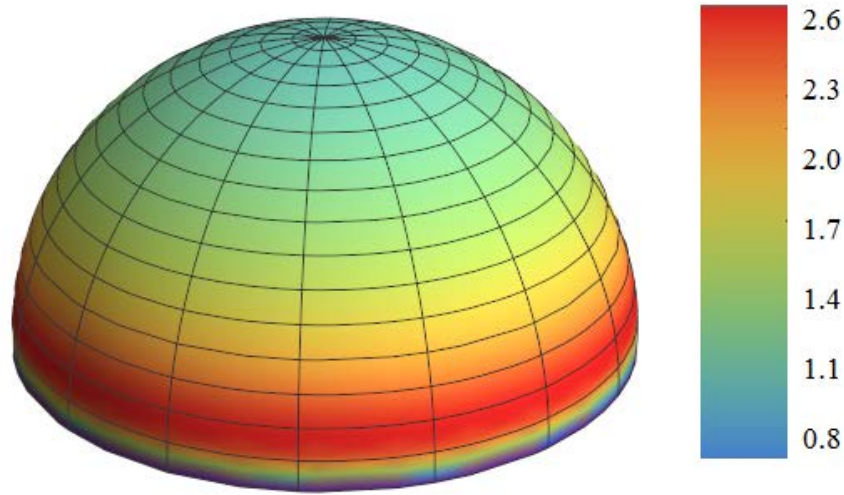

Fig. 3: NSB distribution (in false colors) in an isotropic scattering medium calculated after Eq. (10) for $t$=0.2.

The formula for isotropic scattering is identical to that developed earlier for an isolated source of light (i.e. Eq. 10), except for the fact that now the number of light sources can be arbitrary. Instead of $L\left(z, A = A_0 + \frac{\pi}{2}\right)$ at the left-hand-side of Eq. (10) we can now write $L(z, A)$.

## 4. Numerical demonstrations

The model developed in the section above is advantageous not only because of the exceptionally low number of scaling parameters, but also because it allows for describing a wide range of skies smoothly transitioning from one state to another. The sky type is controlled by two parameters, $t$ and $g$, that have theoretical foundation and are defined on a bounded interval. This shortens the time needed for searching the optimum values of $t$ and $g$ that best match real conditions. Due to their dimensionless nature, $t$ and $g$ facilitate working with the model and help to avoid errors from potential misapprehension when using different units by different authors (this is notoriously typical for some photometric, radiometric and physical quantities).

In the numerical demonstrations below we test the effect of the parameter $t$, which is varied from 0.1 to 0.6 aiming to document the crossover from low to high optical attenuation in the atmosphere, while allowing the parameter $g$ to float within its typical range. We have chosen three discrete values, specifically $g$=0.1 (Fig. 4), $g$=0.4 (Fig. 5), $g$=0.7 (Fig. 6). Two light sources are located at azimuth angles $A_1$=120° and $A_2$=200°, taking into account that the relative radiance of the second source compared to the first is $L_2:L_1$ =2:1. For $g$=0.1 the optical properties approach those of a quasi-isotropic medium, thus strengthening the azimuthal symmetry (see Fig. 4). Low values of $t$ still favor some kind of NSB gradation from zenith toward horizon (Fig. 4a), while the NSB decline rate is reduced markedly when transitioning from $t$=0.1 to $t$ =0.6 (Fig. 4d). At the same time the azimuthal uniformity extends to other parts of the sky.

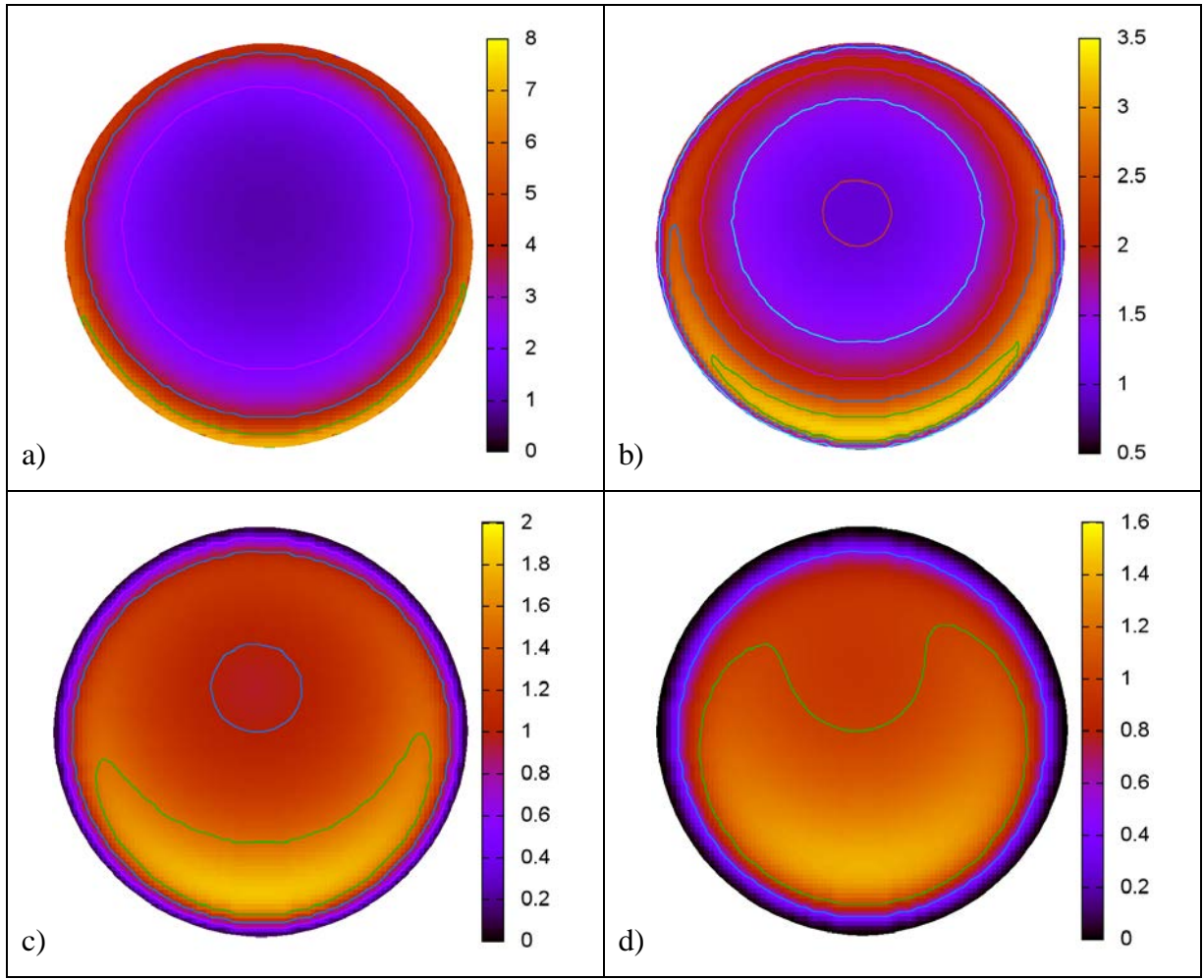

Fig. 4: Zenith-normalized night-sky brightness in false colors computed after Eq. (12) for $g$=0.1 (which promotes nearly isotropic scattering). Two light sources are situated at azimuth angles $A_1$=120° and $A_2$=200°, while their radiances are $L_1$ and $L_2 = 2L_1$, respectively. From

(a) to (d): (a) $t=0.1$, (b) $t=0.2$, (c) $t=0.4$, and (d) $t=0.6$. The lower $t$, the lower the optical attenuation. For each figure the azimuth is measured in clockwise direction (north is at the top). The zenith and horizon are in the center and at the edge of each plot, respectively.

The impact of moderately low-to-average values of $g$ on the zenith-normalized NSB is analyzed in Fig. 5. The value of $g=0.4$ that we have used in the numerical calculations indicates that the scattering medium prefers forward-lobed scattering patterns rather than isotropic scattering functions. Nevertheless, the forward scatter peak is not large enough to cause the sky patches with highest radiances concentrate around the source. Therefore, the sky remains bright not only near the horizon at the position of the light source, but also in adjacent neighborhood areas. Large values of $t$ cause that the atmosphere belongs to a class of polluted media that are expected to asymptotically mimic a cloudy layer. This is why we see a blurred projection of the light source on the sky. The larger $t$, the more pronounced the optical distortion (compare Fig.5c and 5d). The model also suggests that the range of NSB values scales down when the optical attenuation in the atmosphere increases (see e.g. Fig. 5a and 5d).

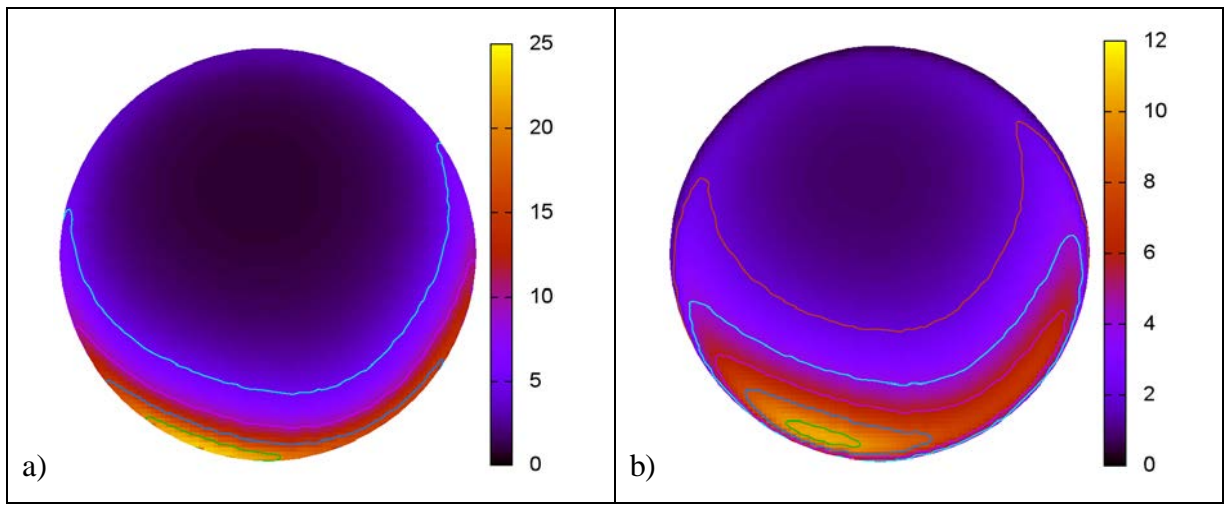

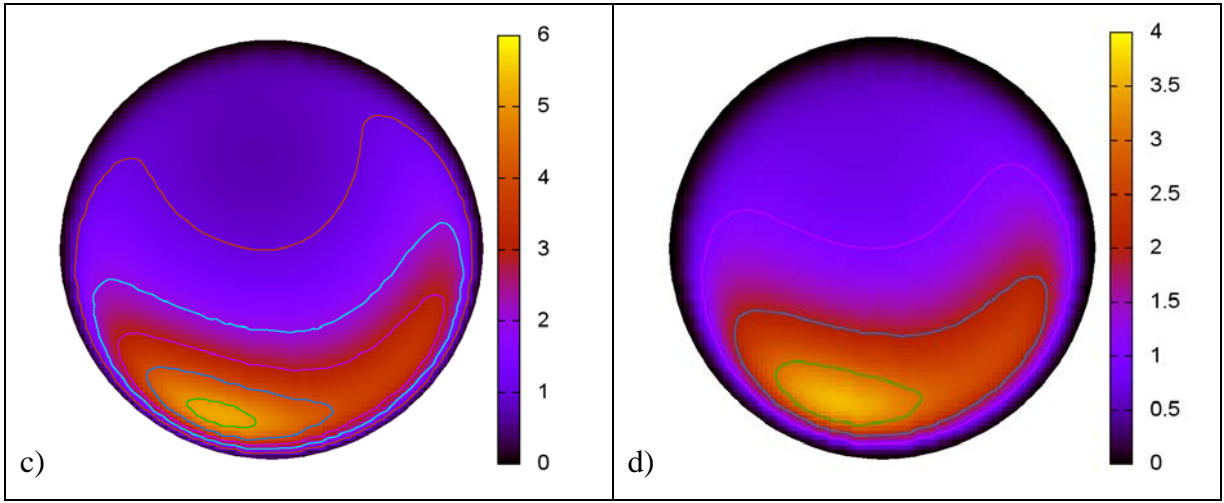

Fig. 5: The same as in Fig.4, but for $g=0.4$ (which promotes nearly isotropic scattering).

At large angular distances from the light source the night sky appears dark when $g$ approaches unity - which is the theoretical limit that rarely or never realizes in nature. However, the side scatter is significantly lowered also for values of $g$ as large as 0.7, which is the reason why the dynamic range of the sky brightness is so high in Fig. 6a. Indeed, the measured NSB usually does not experience such a steep drop in its value, mainly because of the lower limit of natural radiance of the background. By far most of the luminous energy concentrates in a small part of sky, shaped into a bright dome of light over each source. The overall spread of the bright zone to higher elevation angles is larger when the optical attenuation coefficient increases (from Fig. 6a to Fig. 6d). The common feature of the model is that the night sky brightness spans a wide range (approximately two orders of magnitude) when $g$ is above 0.7 and a small range when $g$ is as that for isotropic scattering (compare Figs. 4, 5, and 6).

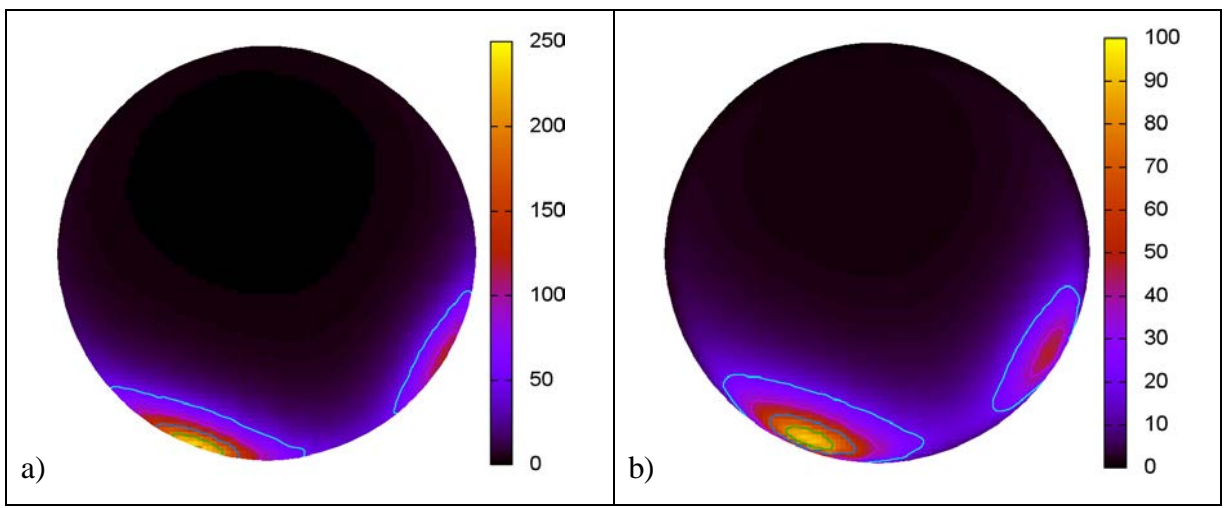

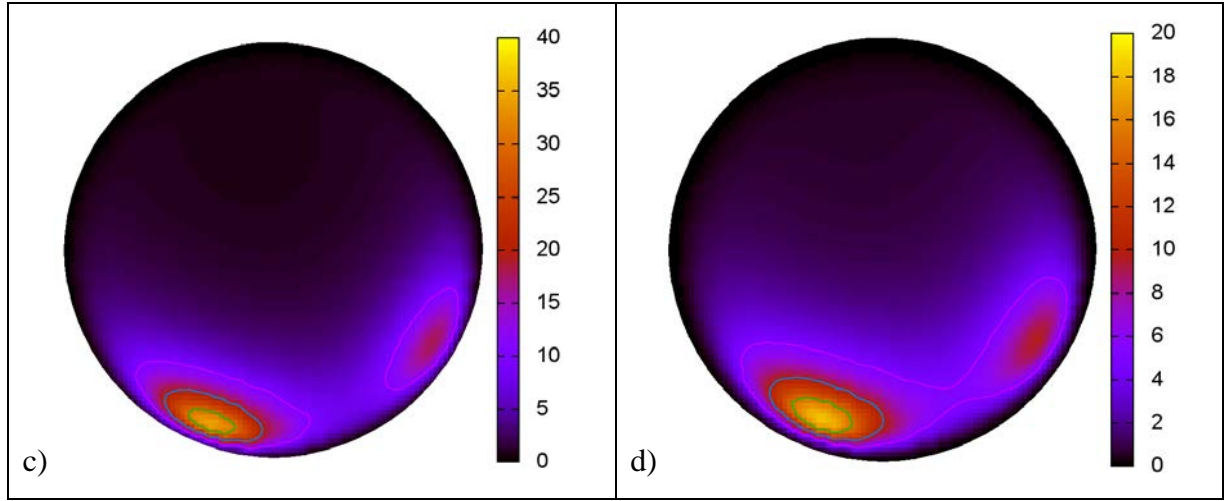

Fig. 6: The same as in Fig.4, but for $g$=0.7 (which promotes strongly anisotropic scattering).

## 5. Experimental validation and retrieval of the parameters that match the observed NSB

To validate the model we have analyzed NSB data taken in a locality whose artificial NSB is mostly due to the emissions from *i*) a single source of light, whereas the parasitic light from other sources is negligible (see Fig. 7), and *ii*) more than two light sources (see Fig. 8). In the first case, the sky brightness distribution on a moonless night partly suffers from atmospheric imperfections in form of an unstable mist that extends over different parts of sky and causes additional veiling luminance (see Fig. 7a). Because of this, the NSB was found to fluctuate around a certain mean in each sky element, while the sky brightness at low elevation angles remains less affected than values near the zenith. The theoretical sky brightness distribution that best match the measured data was found by minimizing the functional

$$f_{t,g}^2 = \int_{z,A} \left[ \frac{L_{t,g}^T(z,A)}{L_{t,g}^T(0,0)} \sin z - \frac{L^E(z,A)}{L^E(0,0)} \sin z \right]^2 dz\, dA \quad, \tag{13}$$

where the integration is carried out over the monitored sky elements (ideally the whole sky). In Eq. (13) the ratios in square brackets are the theoretical ($L_{t,g}^T$) and experimental ($L^E$) zenith-normalized NSB, respectively. The values of $t$ and $g$ that minimize the functional $f_{t,g}^2$ are obtained numerically, while the error is described in terms of $\sqrt{f_{t,g}^2/[2\pi]}$. The factor $2\pi$ is due to normalization to the whole sky vault, $\int_0^{\pi/2} \int_0^{2\pi} \sin z\, dz\, dA$. The experimental brightness data not necessarily are provided on a regular grid, therefore the integral in Eq. (13) is replaced by a sum of $N$ algebraic terms. The standard deviation of the sample for the theoretical-vector is then computed as follows

$$\sigma = \sqrt{\frac{\sum_{i=1}^{N}(F_i^T - F_i^E)^2}{N-1}} \quad , \quad (14)$$

with

$$F_i^T = \frac{L_{t,g}^T(z_i, A_i)}{L_{t,g}^T(0,0)} \sin z_i \quad , \quad F_i^E = \frac{L^E(z_i, A_i)}{L^E(0,0)} \sin z_i \quad . \quad (15)$$

It is demonstrated in Fig. 7b that the theoretical NSB qualitatively explains the experimental data. The peak intensity, the sky brightness in the vicinity of the light source, the radiance gradation from zenith towards horizon, as well as the spread of light over the southeast quadrant, they all are reproduced correctly, excepting for the NSB fluctuation due to unstable haze. The overall error is found to be 3.5%, while the optimum values of the scaling parameters are $t$=0.12, $g$=0.43.

Another experiment was aimed to validate the model in a locality with more than one/two isolated light sources (see Fig. 8a). The sky brightness changes over a wider range of values than that in Fig. 7a, thus indicating a higher atmospheric transparency. Also, the area of elevated radiance does not spread over azimuth angles as much as in Fig. 7a, meaning that the scattering function is more prolonged to forward directions, i.e. the value of $g$ would be most probably higher than that we have found before. The sky was not affected by veiling luminance, thus the overall error of the theoretical fit was as low as 1.8% (compare the observed zenith-normalized NSB in Fig. 8a against the theoretical fit in Fig. 8b). The best fit parameters are consistent with what we have expected, i.e. $t$=0.08, $g$=0.6.

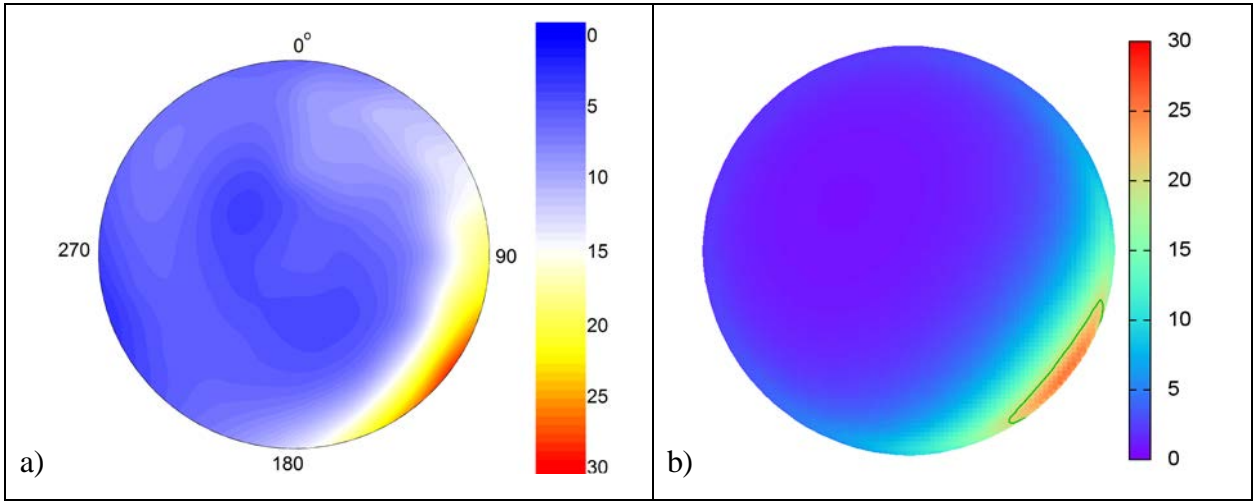

Fig. 7: Experimentally determined zenith-normalized NSB (a) versus the theoretical model obtained as a best match to the measured data (b).

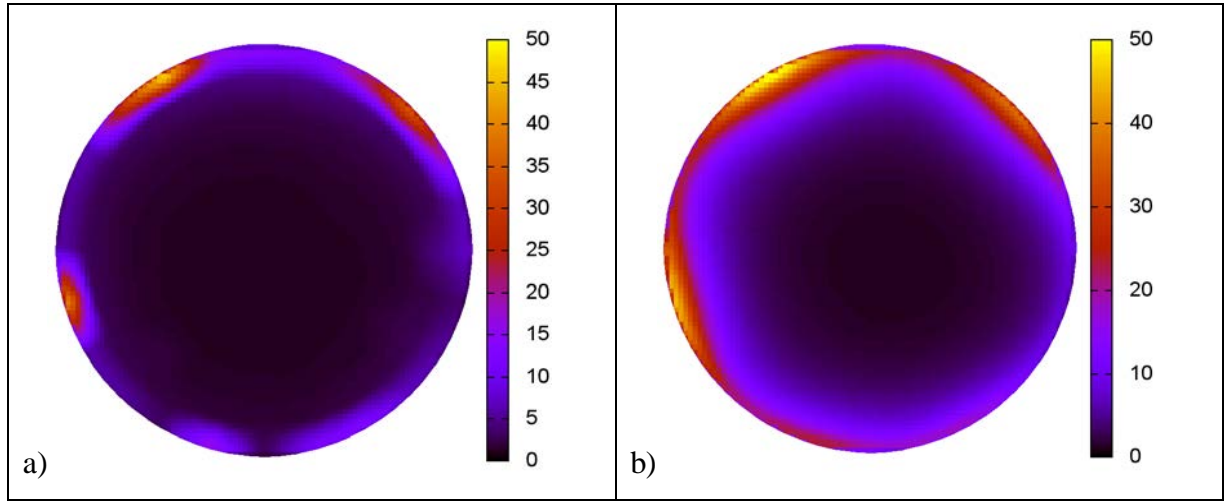

Fig. 8: The same as in Fig. 7, but for another site with more than two important light sources.

## 6. Conclusions

We have shown in this work that the hemispheric night-sky brightness distributions (NSB) produced by artificial light sources can be efficiently described with sufficient accuracy by an analytic family of functions depending only on two indices, $t$ and $g$. The parameter $t$ accounts for the effects of the different atmospheric attenuation, while the parameter $g$ is physically related to the overall degree of asymmetry of the molecular and aerosol scattering processes, and describes how much the scattered photons tend to concentrate near the horizon around the azimuthal position of the light sources.

This model has been heuristically developed from analog results relative to daylight radiance distributions. Some future improvements may incorporate an explicit treatment of the different geometry of both situations, as well as a more detailed account of the variable features of the city emission functions.

The possibility of using two single parameters, plus two additional ones per each relevant light source (namely, their azimuth and relative strength), to represent a wide range of artificial night sky radiance distributions opens interesting possibilities in the field of atmospheric optics and light pollution research. On the one hand, it allows to store all-sky measurement series in an extremely compact form, representing a reduction of order $\sim 10^3$ in the number of required parameters in comparison with the use of Zernike or Legendre expansions, and a reduction of order $\sim 10^6$ in comparison with archiving all pixel values from a typical all-sky radiance image. On the other hand, it facilitates the description of the structure of the night sky brightness at any observing site, and allows a rapid estimation of the general state of the atmosphere from a reduced number of radiance measurements. Finally, given the analytic and explicit form of the elementary radiance distributions corresponding to single light sources, it paves the way for the calculation of world-wide all-sky maps from source radiance data obtained by instruments onboard Earth-orbiting platforms.

**Acknowledgements**

This work was supported by the Slovak Research and Development Agency under contract No: APVV-18-0014, and by Xunta de Galicia/FEDER, grant ED431B 2017/64